\documentclass[sigconf, nonacm]{acmart}
\usepackage{booktabs}
\usepackage{xspace}
\usepackage{enumitem}
\usepackage{graphicx}
\usepackage{overpic}
\usepackage{listings}

\definecolor{javared}{rgb}{0.6,0,0}
\definecolor{javagreen}{rgb}{0.25,0.5,0.35}
\definecolor{javapurple}{rgb}{0.5,0,0.35}
\lstdefinestyle{generatorJava}{
  language=Java,
  basicstyle=\ttfamily\small,
  keywordstyle=\color{javapurple}\bfseries,
  stringstyle=\color{javared},
  commentstyle=\color{javagreen},
  numbers=left,
  numberstyle=\tiny,
  numbersep=6pt,
  xleftmargin=1.8em,
  tabsize=4,
  columns=fullflexible,
  keepspaces=true,
  showspaces=false,
  showstringspaces=false,
  breaklines=true,
  breakatwhitespace=true,
  aboveskip=0pt,
  belowskip=0pt
}
\lstdefinestyle{bugSQL}{
  language=SQL,
  float=tb,
  basicstyle=\ttfamily\footnotesize,
  keywordstyle=\color{black}\bfseries,
  commentstyle=\color{gray},
  numbers=left,
  numberstyle=\scriptsize,
  numbersep=5pt,
  xleftmargin=1.5em,
  columns=fullflexible,
  keepspaces=true,
  showspaces=false,
  showstringspaces=false,
  breaklines=true,
  breakatwhitespace=true
}
\definecolor{taskblue}{RGB}{54,86,114}
\lstdefinestyle{featureTask}{
  style=bugSQL,
  basicstyle=\ttfamily\small,
  language={},
  numbers=none,
  xleftmargin=4pt,
  xrightmargin=4pt,
  frame=single,
  framerule=0.25pt,
  framesep=3pt,
  rulecolor=\color{gray!50},
  backgroundcolor=\color{gray!4},
  morekeywords=[1]{Schema,Target,Request},
  keywordstyle=[1]\color{taskblue}\bfseries,
  morekeywords=[2]{SELECT,FROM,WHERE},
  keywordstyle=[2]\color{black}\bfseries,
  morekeywords=[3]{INT},
  keywordstyle=[3]\color{javagreen}\bfseries,
  aboveskip=3pt,
  belowskip=3pt
}
\makeatletter
\newenvironment{listing}[1][tb]
  {\@float{lstlisting}[#1]}
  {\end@float}
\makeatother

\setcopyright{none}

\newcommand{\SystemName}{\textsc{Gungnir}\xspace}
\newcommand{\SQLancer}{SQLancer\xspace}
\newcommand{\SQLancerPP}{SQLancer++\xspace}
\newif\ifcitationnotes
\citationnotestrue

\newcommand{\ExperimentCPU}{64-core AMD EPYC 7763 CPU at 2.45\,GHz}
\newcommand{\ExperimentMemory}{512\,GB}
\newcommand{\ExperimentOS}{Ubuntu 22.04}
\newcommand{\BaselineRunBudget}{24 hours}
\newcommand{\BaselineCPUs}{4}
\newcommand{\BaselineMemory}{24\,GiB}
\newcommand{\BaselineGeneratorThreads}{1}
\newcommand{\BaselineReplayWorkers}{2}
\newcommand{\BaselineSQLiteVersion}{3.53.4}
\newcommand{\BaselineDuckDBVersion}{1.5.5}
\newcommand{\BaselinePostgreSQLVersion}{18.6}

\newcommand{\BenchmarkTasks}{50}
\newcommand{\FeatureMeasurementSeeds}{3}
\newcommand{\FeatureFuzzSeconds}{600}
\newcommand{\FeatureFuzzThreads}{4}
\newcommand{\FeatureMeasurementCPUs}{4}
\newcommand{\FeatureMeasurementMemory}{12\,GiB}

\newcommand{\MultiWithSuccessrate}{94\%}
\newcommand{\MultiWithoutSuccessrate}{74\%}
\newcommand{\AutonomousWithSuccessrate}{100\%}
\newcommand{\AutonomousWithoutSuccessrate}{4\%}
\newcommand{\MultiWithASTedges}{0.382}
\newcommand{\MultiWithoutASTedges}{0.038}
\newcommand{\AutonomousWithASTedges}{0.004}
\newcommand{\AutonomousWithoutASTedges}{0.155}

\newcommand{\SQLiteLOC}{221}
\newcommand{\DuckDBLOC}{700}
\newcommand{\PostgresLOC}{1,077}

\newcommand{\SQLiteBugs}{18}
\newcommand{\SQLiteFixed}{18}
\newcommand{\SQLiteLogic}{17}
\newcommand{\SQLiteOther}{1}
\newcommand{\DuckDBBugs}{22}
\newcommand{\DuckDBFixed}{14}
\newcommand{\DuckDBLogic}{12}
\newcommand{\DuckDBOther}{10}
\newcommand{\PostgresBugs}{4}
\newcommand{\PostgresFixed}{2}
\newcommand{\PostgresLogic}{3}
\newcommand{\PostgresOther}{1}
\newcommand{\CampaignReports}{45}
\newcommand{\CampaignBugs}{44}
\newcommand{\CampaignFixed}{34}
\newcommand{\CampaignLogic}{32}
\newcommand{\CampaignOther}{12}
\newcommand{\CampaignIntended}{1}

\newcommand{\BaselineSQLiteGungnirLine}{72.18}
\newcommand{\BaselineSQLiteGungnirBranch}{58.60}
\newcommand{\BaselineSQLiteShqvelLine}{39.71}
\newcommand{\BaselineSQLiteShqvelBranch}{31.20}
\newcommand{\BaselineSQLiteSqlancerPPLine}{35.89}
\newcommand{\BaselineSQLiteSqlancerPPBranch}{27.43}
\newcommand{\BaselineSQLiteSqlancerLine}{57.44}
\newcommand{\BaselineSQLiteSqlancerBranch}{45.63}
\newcommand{\BaselineDuckDBGungnirLine}{38.19}
\newcommand{\BaselineDuckDBGungnirBranch}{26.66}
\newcommand{\BaselineDuckDBShqvelLine}{32.98}
\newcommand{\BaselineDuckDBShqvelBranch}{26.38}
\newcommand{\BaselineDuckDBSqlancerPPLine}{26.18}
\newcommand{\BaselineDuckDBSqlancerPPBranch}{17.98}
\newcommand{\BaselineDuckDBSqlancerLine}{27.86}
\newcommand{\BaselineDuckDBSqlancerBranch}{20.54}
\newcommand{\BaselinePostgreSQLGungnirLine}{40.55}
\newcommand{\BaselinePostgreSQLGungnirBranch}{29.91}
\newcommand{\BaselinePostgreSQLShqvelLine}{27.11}
\newcommand{\BaselinePostgreSQLShqvelBranch}{19.96}
\newcommand{\BaselinePostgreSQLSqlancerPPLine}{23.11}
\newcommand{\BaselinePostgreSQLSqlancerPPBranch}{16.49}
\newcommand{\BaselinePostgreSQLSqlancerLine}{30.97}
\newcommand{\BaselinePostgreSQLSqlancerBranch}{22.81}
\newcommand{\BaselineSQLiteBranchGain}{31.17}
\newcommand{\BaselineDuckDBBranchGain}{8.68}
\newcommand{\BaselinePostgreSQLBranchGain}{13.42}

\newcommand{\DuckdbPracticalityLLM}{\texttt{deepseek-flash}}
\newcommand{\DuckdbPracticalityBugTarget}{20}
\newcommand{\DuckdbPracticalityBudgetHours}{24}
\newcommand{\DuckdbPracticalityFuzzThreads}{6--8}
\newcommand{\DuckdbPracticalityAcceptedGoals}{81}
\newcommand{\DuckdbPracticalityAcceptedModifications}{92}
\newcommand{\DuckdbPracticalityTestedRevisions}{82}
\newcommand{\DuckdbPracticalitySegments}{90}
\newcommand{\DuckdbPracticalityStatementsMillion}{29.49}
\newcommand{\DuckdbPracticalityTestingHours}{23.99}
\newcommand{\DuckdbPracticalityBugReports}{5}
\newcommand{\DuckdbPracticalityElapsedHours}{24.29}
\newcommand{\DuckdbPracticalityTotalTokensMillion}{1,773.16}
\newcommand{\DuckdbPracticalityCostUSD}{20.85}

\newcommand{\TestSuiteSQLiteLine}{95.38}
\newcommand{\TestSuiteSQLiteBranch}{87.22}
\newcommand{\TestSuiteDuckDBLine}{61.91}
\newcommand{\TestSuiteDuckDBBranch}{40.83}
\newcommand{\TestSuitePostgreSQLLine}{64.37}
\newcommand{\TestSuitePostgreSQLBranch}{50.61}

\AtBeginDocument{%
  }

\begin{document}

\title{Agentic Evolution of SQL Test Generators}

\author{Suyang Zhong}
\affiliation{%
  \institution{National University of Singapore}
  \city{Singapore}
  \country{Singapore}}
\email{suyang@u.nus.edu}
\author{Manuel Rigger}
\affiliation{%
  \institution{National University of Singapore}
  \city{Singapore}
  \country{Singapore}}
\email{rigger@nus.edu.sg}

\begin{abstract}
Automated testing has uncovered many logic bugs in Database Management Systems (DBMSs).
Existing tools are typically limited to a few SQL dialects and a subset of these dialects' features.
Coding agents based on Large Language Models (LLMs) could encode the semantics of additional SQL features, which would then require no additional LLM calls during these tools' run time, thus allowing high efficiency and low cost.
However, a key challenge is to ensure the correctness of the added features in the presence of potential hallucinations.
In this paper, we propose \SystemName, an approach for evolving DBMS test generators using coding agents.
We use \emph{witness replay} to validate generator modifications by replaying the choices needed to generate a target statement.
A \emph{generation witness} is responsible for recording the database setup and semantic choices needed to generate the target statement.
A validator checks the witness by replaying these choices through the generator and checking that it reproduces the target statement.
We tested \SystemName on SQLite, DuckDB, and PostgreSQL and found \CampaignBugs{} bugs confirmed by developers, of which \CampaignFixed{} have been fixed.
Beyond DBMSs, witness replay could generalize to other systems for guiding agents in extending test generators.
\end{abstract}



\maketitle

\newif\ifcomments
\commentstrue
\ifcomments
    \providecommand{\suyang}[1]{{\color{brown}{suyang: #1 }}}
\else
    \providecommand{\suyang}[1]{}
\fi

\section{Introduction}

Database Management Systems (DBMSs) are complex software systems whose correctness affects downstream applications.
Besides crashes and performance issues, DBMSs can contain \emph{logic bugs}, which cause queries to return incorrect results without raising an error.
Automated testing approaches detect these bugs by generating SQL statements and checking their results with a \emph{test oracle}~\cite{rigger2020finding,rigger2020detecting}.
These approaches have uncovered hundreds of previously unknown bugs in widely used DBMSs~\cite{zhong2025scaling,zhong2025shqvel}.

The effectiveness of a test oracle depends on which SQL features the generated statements contain.
Mutation-based tools derive inputs from existing statements, and their scope depends on the seed corpus and available mutation operators~\cite{zhong2020squirrel}.
Generation-based tools create schemas, populate tables, and generate queries from predefined rules~\cite{rigger2020detecting,rigger2020finding,rigger2020testing}.
Once implemented, a generator can repeatedly test its supported features at high throughput.
The challenge of supporting the multitude of SQL features within a generator is further exacerbated given that most DBMSs implement a distinct SQL dialect.
\SQLancerPP supports an adaptive generator that adjusts generation probabilities during run-time to fit different dialects, but requires manual source modifications to support new SQL features~\cite{zhong2025scaling}.
Listing~\ref{lst:sqlite-evolution-bug} shows a reduced test case that exposes a logic bug in SQLite, which \SQLancerPP fails to find due to lack of support for row value \texttt{IN} predicates with aggregate subqueries.
The existing test oracle can detect this bug once the generator supports the required predicate.
Automating this task would allow the generator to incorporate new SQL features as DBMSs evolve.
However, adding such support requires understanding both the generator implementation and the SQL features of the target DBMS.

\begin{lstlisting}[
  language=SQL,
  float=tb,
  caption={A SQLite logic bug exposed after generator evolution. The query should return \texttt{-1}, but returns no rows.},
  label={lst:sqlite-evolution-bug},
  numbers=left,
  numberstyle=\scriptsize,
  basicstyle=\ttfamily\footnotesize,
  keywordstyle=\bfseries,
  commentstyle=\color{gray},
  showstringspaces=false,
  breaklines=true,
  xleftmargin=1em]
CREATE TABLE t0(c1 INTEGER, UNIQUE(c1));
INSERT INTO t0 VALUES (5), (-1);

SELECT c1 FROM t0 WHERE (c1, c1) IN
  (SELECT c.c1, min(c.c1) FROM t0 AS c);
-- Expected: -1; Actual: no rows
\end{lstlisting}
Recent work uses Large Language Models (LLMs) within predefined workflows to automatically improve DBMS testing tools.
For example, ShQveL uses LLMs to add validated SQL fragments to the generator, while Argus uses LLMs to propose test oracles that are independently verified~\cite{zhong2025shqvel,mang2026argus}.
However, the SQL features supported by these workflows depend on their generation and verification mechanisms.
ShQveL fills predefined sketches, while Argus verifies query relations within the SQL features supported by its underlying solver and base generator.
Both generators lack support for generating the row value \texttt{IN} comparison with a subquery to trigger the bug in Listing~\ref{lst:sqlite-evolution-bug}.
Extending these mechanisms to support additional SQL features still requires modifications to the test generator.

Increasing LLM capability opens new opportunities for greater autonomy than in previous work~\cite{luo2026data}.
Coding agents can inspect software repositories and implement modifications across several source files~\cite{yang2024swe,zhang2024autocoderover,wang2025openhands}.
These agents could implement the state handling and generation rules needed for a new SQL feature.
However, a key challenge is to deterministically \emph{validate} whether the modified generator can generate statements containing the requested feature.
Current feedback signals provide insufficient evidence for the agents to determine when a certain feature can be generated.
First, checking for a successful build provides insufficient evidence that the generator can generate requested features during run-time.
Second, it is difficult to validate whether the generator can generate a requested feature, since, due to the inherent stochasticity of generators, the feature would not necessarily be generated within a fixed timeframe.
Third, coverage feedback can guide LLM-driven generator evolution~\cite{chen2025elfuzz}, but an increase in coverage does not show that a requested statement can be generated.

In this paper, we propose \SystemName, a system for evolving DBMS test generators with coding agents.
Our core insight is to require the coding agent to provide a so-called \emph{generation witness} alongside each generator modification, providing concrete evidence of how the modified generator can generate a target statement.
We realize this witness mechanism through a common guidance interface that records the \emph{semantic choices} instrumented in the generator.
The input to the witness mechanism is a target SQL statement---which can contain one or more new SQL features.
A generation witness records the database setup and semantic choices needed to generate this target.
The coding agent is responsible for implementing the requested feature and identifying the choices needed to generate the target; the validator is responsible for checking this evidence by replaying the choices through the modified generator, comparing the complete generated statement with the target statement, and running the required executable check before the modification can be accepted for further testing.
For subsequent modifications, we require the coding agent to preserve previously witnessed behavior, which the validator checks by replaying the historical witnesses.

We implement \SystemName as a multi-agent system (MAS) that evolves \SQLancerPP, separating the analysis of missing DBMS features from their integration into the test generator.
An analyst agent derives target statements from DBMS source code, tests, and bug reports, while a coding agent modifies the generator using witness feedback.
A triage agent investigates failures found during DBMS testing.
The testing results and triage findings guide the analyst agent's selection of further targets, forming a controlled evolution loop.
After evolution, no more calls to the LLM or witness mechanism are required.

We evaluated \SystemName on SQLite, DuckDB, and PostgreSQL and reported \CampaignBugs{} bugs confirmed by developers, of which \CampaignFixed{} have been fixed.
The evolved generators achieved higher line and branch coverage than the generator baselines on all three DBMSs in the reported runs.
In a separate DuckDB case study fully automated by agents---starting from a single prompt and ending with bug reports---\SystemName completed \DuckdbPracticalityAcceptedGoals{} feature goals and generated \DuckdbPracticalityBugReports{} bug reports confirmed by developers, with a cost of US\$\DuckdbPracticalityCostUSD{}.

In summary, this paper makes the following contributions:

\begin{itemize}[leftmargin=*]
  \item We propose \emph{witness replay}, which checks whether a modified generator can generate target statements and supplies feedback for further source modifications.
  \item We implement \SystemName, a system that evolves \SQLancerPP with coding agents and preserves the resulting features for DBMS testing.
  \item We evaluate \SystemName on SQLite, DuckDB, and PostgreSQL, reporting \CampaignBugs{} bugs confirmed by developers, including \CampaignFixed{} fixed bugs.
\end{itemize}

\section{Background and Motivation}
\label{sec:background}

We first introduce SQL generators and test oracles, then discuss how LLMs can help generate SQL test cases.

\subsection{SQL Test Generation}

\paragraph{Test generator.}
SQL generators construct test cases from scratch based on predefined rules~\cite{rigger2020testing,rigger2020finding,zhong2025scaling}.
The rules define how to create a database schema, populate tables, and generate queries following target DBMS semantics.
First, generators track the database state to satisfy dependencies among statements.
For example, a table must exist before an insertion, and a query can reference only columns available in the current schema.
Second, the generator needs to generate statements that are syntactically and semantically valid for the target DBMS.
For example, it selects columns and operations with compatible types when constructing expressions.
SQL dialects add further constraints on which features a DBMS accepts~\cite{zhong2024quality,zhong2025scaling}.
For example, SQLancer supports 22 DBMS-specific generators with an average of 3.1K lines of code (LOC) for generation and supporting components.
Supporting a new feature requires modifications to schema handling, expression generation, and SQL rendering.
Third, the generator needs to satisfy the test oracle's syntactic and semantic constraints.
These requirements extend to generators that are modified by an LLM, such as in our work.

\paragraph{Test oracle.}
A test oracle determines the expected result of a test case.
It allows the testing tool to check whether a randomly generated SQL statement returns the expected result.
Ternary Logic Partitioning (TLP) is a widely used oracle for logic bug detection~\cite{rigger2020finding}.
TLP divides the rows of one query according to whether a random predicate evaluates to true, false, or NULL, then compares the union of these partitions with the original result.
While \SystemName relies on TLP in its implementation, it can adapt any of the existing test oracles.
We deliberately avoid using LLMs for generating test oracles, since an incorrect one---for example, caused by hallucinations---would result in false alarms that would require developer intervention.

\paragraph{SQLancer++.}
\SQLancerPP reduces the effort of supporting different SQL dialects by generalizing generation rules across DBMSs~\cite{zhong2025scaling}.
Its adaptive generator uses execution feedback to adjust probabilities of generating implemented SQL features; however, generating additional features requires new rules.
For example, the SQLite case in Listing~\ref{lst:sqlite-evolution-bug} required a row value representation and generation logic that connected it to an aggregate subquery.
Adding this feature required modifications to the predicate generation of \SQLancerPP.

Listing~\ref{lst:generator-base} shows the existing rule for an \texttt{IN} predicate with a list of expressions.
The predicate generator calls \texttt{booleanExpr}, which first selects an enabled Boolean expression form.
For \texttt{IN\_LIST}, it selects a type and generates the left operand and list elements with that type.
Adding a new kind of Boolean expression requires adding its construction rule and a branch that calls it.
For the example in Listing~\ref{lst:sqlite-evolution-bug}, this means generating a row value on the left and an aggregate subquery on the right of \texttt{IN}.

\begin{listing}
  \caption{Boolean expression generation in \SQLancerPP. Other operations are omitted.}
  \label{lst:generator-base}
\begin{lstlisting}[style=generatorJava,escapeinside={(*@}{@*)}]
Expr booleanExpr(Context c) {
  Form f = choose(c.booleanForms); (*@\label{line:base-choice}@*)
  if (f == IN_LIST) {
    Type t = choose(c.types);
    Expr x = expression(c, t);
    List<Expr> xs = expressions(c, t, smallInt()+1);
    return in(x, xs);
  }
  return otherBoolean(c, f);
}
\end{lstlisting}
\end{listing}

\subsection{LLMs for Test Generation}

\paragraph{LLM-generated SQL}
A plausible way to automate DBMS testing is to let coding agents \emph{directly} generate SQL test cases and execute them to then investigate potential bugs, potentially also allowing them to inspect the DBMS' source code.
However, this direct approach faces three challenges.
First, the agent can infer an incorrect expected result and report correct behavior as a logic bug.
This difficulty also occurs in general software testing: RepoAudit reports false positives in code auditing~\cite{guo2025repoaudit}, and agentic property-based testing documents agents mistaking intended behavior for bugs~\cite{maaz2025agenticpbt}.
Second, generating test cases requires many LLM calls unless the generation logic is saved for reuse, adding inference cost and latency to continued testing.
Third, the search space of agents can be biased toward familiar bug patterns learned during pretraining, such as incorrect handling of boundary values.
These patterns can guide useful bug findings, but may also concentrate testing on similar cases and leave unfamiliar feature combinations unexplored.
These difficulties motivate using agents to extend a reusable SQL generator.

\paragraph{LLM-based SQL generators.}
Retaining \emph{validated} LLM knowledge in a generator allows repeated testing without further model calls and avoids the false alarms that arise from relying on LLMs to judge correctness.
ShQveL fills SQL sketches with LLMs and saves validated fragments for subsequent test generation~\cite{zhong2025shqvel}.
It checks filled sketches by execution and uses existing test oracles to detect logic bugs.
Argus discovers pairs of abstract queries, verifies their equivalence with a SQL equivalence solver, and instantiates them with reusable SQL snippets~\cite{mang2026argus}.
Its soundness depends on the verified equivalence and the constraints imposed on snippet instantiation.
Both approaches validate LLM outputs before using them for testing instead of relying on the LLM to judge correctness.

Both ShQveL and Argus are constrained by the generator's existing capabilities.
ShQveL can learn new fragments, but the locations of sketch holes and the generator logic that uses them remain manually implemented.
Specifically, its expression sketches fill developer-selected operator or function holes; they cannot extend the generator with new syntactic structures, for example, the row value \texttt{IN} comparison in Listing~\ref{lst:sqlite-evolution-bug}.
Argus is further limited by the SQL features supported by its equivalence solver, and a query relation that the solver cannot prove cannot become a verified oracle~\cite{mang2026argus}.
For example, it cannot directly prove SQLite's bare-column aggregate semantics in Listing~\ref{lst:sqlite-evolution-bug}.
Although snippets can introduce additional features during instantiation after suitable extensions, they require substantial LLM effort to build a snippet corpus.

\section{Approach}
\label{sec:approach}

\begin{figure*}[t]
  \centering\includegraphics[width=\textwidth]{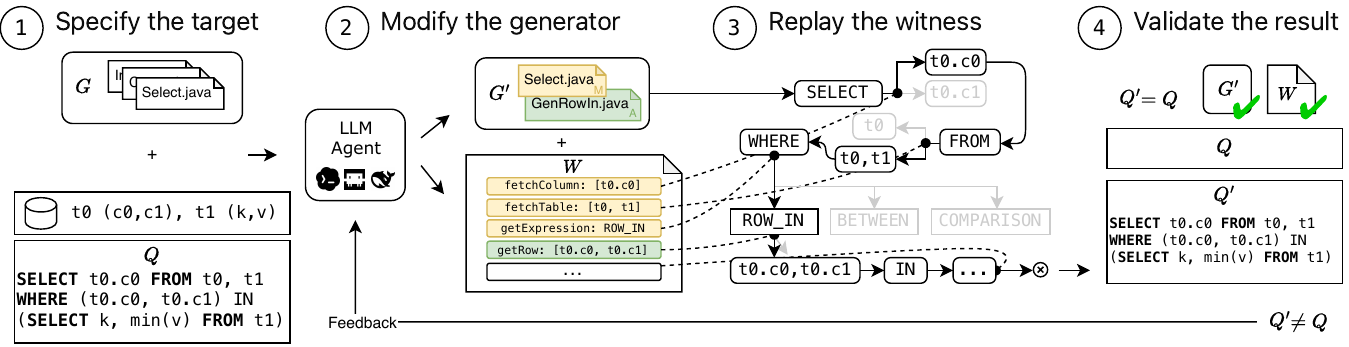}
  \caption{Witness replay for a row value \texttt{IN} target. The modified generator $G'$ uses witness $W$ to generate $Q'$, which is compared with target $Q$. File badges \texttt{M} and \texttt{A} mark modified and added source files. Black branches show selected alternatives, gray branches show unselected alternatives, and dashed lines bind witness choices to generation decisions.}
  \Description{A diagram with four stages. The first provides the current generator, a target row value IN statement, and its database setup. The second shows an LLM coding agent modifying the generator and constructing a generation witness. The third applies witness choices to select the projection, tables, and row value IN predicate through the modified generator. The fourth compares the generated statement with the target and returns mismatches to the coding agent.}
  \label{fig:witness-mechanism}
\end{figure*}

In this paper, we propose \SystemName, which tackles the problem of reliably building and continually improving automated DBMS testing tools for finding logic bugs.
The core mechanism of \SystemName is \emph{witness replay}, which checks whether a modified generator can generate a requested SQL statement.
We assign a coding agent the responsibility of modifying the generator to support given target SQL statements---which provide concrete examples of feature usage---and providing executable evidence of that support.
A \emph{generation witness} records the database setup and semantic choices needed to generate a target statement.
An external validator checks whether the agent has fulfilled this responsibility by replaying each witness through the modified generator, comparing the generated statement with its target.
Failures guide further modifications to the generator and its witness, and the witnesses are saved for regression testing of future generator modifications.
Figure~\ref{fig:witness-mechanism} illustrates these steps for a row value \texttt{IN} predicate whose right side is an aggregate subquery. We next detail these steps.

\subsection{Specifying the Target}

In Step~\textcircled{1}, our approach expects a target SQL statement $Q$ containing the feature to be added.
The target illustrates the feature for the coding agent and specifies the required behavior for the validator.
Targets can be derived automatically by agents from DBMS documentation, source code, regression tests, bug reports, or prior testing results.
In principle, they can also be supplied manually.
Targets should satisfy several constraints, including being valid SQL statements and meeting the input requirements of the test oracle used for validation.
We explain how these constraints guide the analyst agent in synthesizing targets in Section~\ref{sec:goal-constraints}.
During target synthesis, the existing test oracle checks the proposed test case on the selected DBMS before it becomes a target for the coding agent.
For example, the target $Q$ in Figure~\ref{fig:witness-mechanism} selects \texttt{t0.c0} from \texttt{t0} and \texttt{t1}, with a predicate comparing the row \texttt{(t0.c0, t0.c1)} with the result of \texttt{SELECT k, min(v) FROM t1}.

\subsection{Modifying the Generator}
In Step~\textcircled{2}, we require the coding agent to modify the current generator $G$ to generate the target statement and construct a generation witness that demonstrates this behavior.
We denote the modified generator by $G'$.

\paragraph{Source modifications.}
We call the coding agent to modify source code to add new features for generating the target $Q$.
The agent adds \texttt{ROW\_IN} to the Boolean expression forms in Listing~\ref{lst:generator-base}.
Listing~\ref{lst:generator-extension} shows the new branch and helper.
The context \texttt{c} provides the schema tables through \texttt{c.tables} and the outer query's columns through \texttt{c.columns}.
The generator selects \texttt{ROW\_IN} through the same choice as \texttt{IN\_LIST} and the other Boolean expressions.
This enables the generator to generate the target $Q$.
During execution, the generator may pick corresponding choices for the new \texttt{ROW\_IN} branch, its tables and columns, and the \texttt{MIN} aggregate.
However, agents might make mistakes when performing changes.
For example, the agent may add the new predicate without making it reachable from the query generator, or may simply call the wrong function that generates only constants but not column references.
We thus need to validate that the modified generator can generate the target.

\begin{listing}
  \caption{Generator modification for Step~\textcircled{2}. Yellow shading marks the added \texttt{ROW\_IN} branch, and green shading marks its helper.}
  \label{lst:generator-extension}
\begin{lstlisting}[style=generatorJava,escapeinside={(*@}{@*)},firstline=1,lastline=5]
Expr booleanExpr(Context c) {
  Form f = choose(c.booleanForms); (*@\label{line:extension-form}@*)
  if (f == IN_LIST) {
    ...
  }
\end{lstlisting}
\begin{lstlisting}[style=generatorJava,escapeinside={(*@}{@*)},firstline=6,lastline=7,firstnumber=6,backgroundcolor=\color{yellow!10}]
  if (f == ROW_IN) (*@\label{line:extension-branch}@*)
    return rowIn(c);
\end{lstlisting}
\begin{lstlisting}[style=generatorJava,escapeinside={(*@}{@*)},firstline=8,lastline=10,firstnumber=8]
  return otherBoolean(c, f);
}
\end{lstlisting}
\begin{lstlisting}[style=generatorJava,escapeinside={(*@}{@*)},firstline=11,firstnumber=11,backgroundcolor=\color{green!10}]
Expr rowIn(Context c) {
  Table t = choose(c.tables); (*@\label{line:extension-table}@*)
  Agg op = choose(MIN, MAX); (*@\label{line:extension-aggregate}@*)
  Column v = choose(t.columns); (*@\label{line:extension-value}@*)
  Expr agg = call(op, v); (*@\label{line:extension-call}@*)
  Column k = choose(t.columns); (*@\label{line:extension-key}@*)
  Query q = select(t, k, agg); (*@\label{line:extension-select}@*)
  Column a = choose(compatible(c.columns, k)); (*@\label{line:extension-first-column}@*)
  Column b = choose(compatible(c.columns, v)); (*@\label{line:extension-second-column}@*)
  return in(row(a, b), q); (*@\label{line:extension-return}@*)
}
\end{lstlisting}
\end{listing}

\paragraph{Generator model.}
To describe the choices needed for such an execution, we model SQL generation as a sequence of semantic decisions.
This model follows the interpretation of a random generator as a parser of choice sequences~\cite{goldstein2022parsing}.
Each $d_i$ identifies a decision site together with its occurrence number during generation.
For example, a column-selection call inside a loop may run twice to construct a two-column expression; its first execution can select \texttt{t0.c0} and its second \texttt{t0.c1}.
The source location and occurrence number distinguish these decisions in the witness.
The state $x_i$ contains the current schema, expected type, partial Abstract Syntax Tree (AST), and local generation context before decision site $d_i$.
We write $D_{G'}(d_i,x_i)$ for the alternatives allowed by the generator and its fixed configuration at decision site $d_i$ in state $x_i$, and $T_{G'}$ for the generator's state transition after selecting an alternative.
A semantic decision can summarize internal random draws and sampling retries; $T_{G'}$ describes the construction step that uses the selected value.
A selection policy $\pi$ returns an alternative from $D_{G'}(d_i,x_i)$.
During random generation, the policy $\pi_0$ uses pseudorandom draws initialized by a seed to select alternatives.
Thus, $G'$ chooses
\[
  c_i=\pi_0(d_i,x_i)\in D_{G'}(d_i,x_i), \qquad
  x_{i+1}=T_{G'}(x_i,d_i,c_i).
\]
The policy determines which alternative is selected; $T_{G'}$ determines how the generator uses it to construct SQL.
A decision can select a table, column, expression form, operator, constant, or statement generation strategy.
In Listing~\ref{lst:generator-extension}, each \texttt{choose} call corresponds to a decision site, and its arguments determine $D_{G'}(d_i,x_i)$.
For example, at the aggregate choice on line~\ref{line:extension-aggregate}, $D_{G'}(d_i,x_i)=\{\texttt{MIN},\texttt{MAX}\}$; selecting $c_i=\texttt{MIN}$ stores that aggregate in \texttt{op} for the subsequent construction step.
Domains can also depend on the current context. For example, \texttt{choose(c.tables)} selects from the available tables, while \texttt{choose(compatible(c.columns, k))} selects columns compatible with \texttt{k}.
Although the listing shows choices from collections, the model also covers random integers drawn from a range $[\ell,u)$, with domain $\{z\in\mathbb{Z}\mid \ell\leq z<u\}$.
Replay checks range bounds directly without enumerating the integers.
Like Zest, which exposes generator parameters to guide structured test generation~\cite{padhye2019zest}, we make generator choices explicit.
Semantic names such as \texttt{MIN} and \texttt{t0.c1} let the agent identify the decisions needed for the target.

\paragraph{Witness construction.}
We require the coding agent to provide a witness that can be externally validated based on the generator model.
To this end, during source modification, the agent identifies the choices needed to generate the target and records them in a witness.
An executable setup $F$ creates the schema and data required by a target.
A generation witness $W=(F,A)$ pairs this setup with an assignment $A$ from decision sites to semantic values.
The setup $F$ initializes the generation state $x_0$, which contains the table names (\emph{e.g.}, \texttt{t0, t1}) and the column names (\emph{e.g.}, \texttt{c0, c1}) for the target.
The assignment $A(d_i)$ selects the alternative used by $T_{G'}$ whenever $d_i$ is assigned.
A valid assignment thus satisfies $A(d_i)\in D_{G'}(d_i,x_i)$ at each assigned decision reached during execution.
During replay, the validator checks each assigned value against the alternatives allowed by the generator in the current state $x_i$.
The witness specifies which alternatives to select through the generator $G'$, which then executes the corresponding construction steps to generate SQL.
Selecting an existing alternative only requires changing the witness assignment.
If a required alternative, such as \texttt{ROW\_IN}, is unavailable, the agent should modify the generator to support it.
Further source modifications can change decision sites and require corresponding changes to $A$.
The agent submits both $G'$ and $W$ for replay and validation against the fixed target $Q$.

For the running example in Listing~\ref{lst:generator-extension}, $A$ selects \texttt{ROW\_IN} at line~\ref{line:extension-form} and \texttt{t1} at line~\ref{line:extension-table}.
It selects \texttt{MIN} at line~\ref{line:extension-aggregate}, then \texttt{t1.v} and \texttt{t1.k} at lines~\ref{line:extension-value} and~\ref{line:extension-key}, respectively.
Finally, it selects \texttt{t0.c0} and \texttt{t0.c1} for the row value at lines~\ref{line:extension-first-column} and~\ref{line:extension-second-column}.
Together with the choices of \texttt{t0.c0} for the outer projection and \texttt{t0} and \texttt{t1} for the outer \texttt{FROM} clause, these assignments direct $G'$ to construct \texttt{(t0.c0, t0.c1) IN (SELECT k, min(v) FROM t1)}.

\subsection{Replaying the Witness}
\label{sec:witness-replay}

In Step~\textcircled{3}, the validator checks whether the coding agent has correctly implemented the feature in the generator so that it can generate the target $Q$.
It does so by executing the setup $F$ and running $G'$ under the policy $\pi_A$ specified by the assignment $A$:
\[
  G'(F;\pi_A)=Q'.
\]
Here, $Q'$ denotes the generated statement designated for comparison with the target statement $Q$.
Replay requires no LLM calls or human inspection.
For the row value target, $F$ creates tables \texttt{t0(c0 INT, c1 INT)} and \texttt{t1(k INT, v INT)} and initializes the schema in $x_0$.

\paragraph{Replay policy.}
Replay changes the selection policy while preserving the generator's construction transitions $T_{G'}$.
To make the generator replayable, we instrument its choice operations, represented by \texttt{choose} in Listing~\ref{lst:generator-extension}.
At each call, the instrumentation identifies the decision site and its occurrence, obtains the alternatives allowed in the current state, and consults $A$.
It returns the assigned value when one is present and otherwise invokes the original random selection procedure.
Thus, during witness construction and new-target validation, the replay policy is
\[
  \pi_A(d_i,x_i)=
  \begin{cases}
    A(d_i), & \text{if } A(d_i) \text{ is defined},\\
    \pi_0(d_i,x_i), & \text{otherwise}.
  \end{cases}
\]

\paragraph{SQL construction.}
Replay checks $A(d_i)\in D_{G'}(d_i,x_i)$ at each assigned decision site before applying the generator's transition $T_{G'}$.
For the aggregate decision site $d_a$ on line~\ref{line:extension-aggregate} of Listing~\ref{lst:generator-extension}, $A(d_a)=\texttt{MIN}$ is valid because its domain contains \texttt{MIN} and \texttt{MAX}; assigning \texttt{COUNT} would fail this check.
The transitions use these choices to construct the aggregate expression, its enclosing subquery, and the row value comparison in Figure~\ref{fig:witness-mechanism}.
Replay records the visited decision sites and selected values and fails if an assigned site is never reached.

\subsection{Validating the Result}
In Step~\textcircled{4}, the validator checks both the requested behavior and previously witnessed behavior before accepting $G'$.

\paragraph{New target validation.}
The generation requirement is that replay uses all assigned choices successfully and generates the fixed target by exact string equality $Q'=Q$.
This comparison checks whether the modified generator can generate it.
Starting from $x_0$, each replayed decision selects $c_i\in D_{G'}(d_i,x_i)$ and applies the generator's transition $T_{G'}$.
By induction over these decisions, replay follows a path through the modified generator's own transitions.
Together with $Q'=Q$, this shows that $G'$ can generate the target from $F$ under the witness assignment.
As a limitation, a passing witness cannot show whether or how frequently $Q$ occurs during random testing, as discussed in Section~\ref{sec:discussion}.

\paragraph{Historical target validation.}
\label{sec:historical-replay}
We maintain a regression test suite $\mathcal{W}$ of accepted generation witnesses to preserve previously witnessed behavior.
For each accepted witness $W=(F,A)$, we record all choices from its successful execution in $A$ and add $W$ to $\mathcal{W}$, together with the expected SQL.
Before accepting a later generator modification $G'$, the validator replays every $W\in\mathcal{W}$ through $G'$ using its setup and recorded choices.
The modification is accepted only if every witness in the suite passes.
The agent may update $A$ when the decision sites change, while the setup and the expected SQL remain fixed.
If source edits within the same file shift site locations, the agent updates those locations automatically.
Otherwise, ambiguous changes require the agent to inspect the affected code and revise the witness.%

\paragraph{Feedback.}
Replay failures provide feedback to the coding agent for modifying the generator and its witness assignments.
Witness validation can fail in three ways.
First, an assigned value can be unavailable: $A(d_i)\notin D_{G'}(d_i,x_i)$ at a visited decision.
Second, the assignments can fail to match the visited decision sites: an assigned site is never reached, or a visited site has no assignment during strict historical replay.
Third, the generated statement can differ from the target statement ($Q'\neq Q$), or historical replay can fail to reproduce the recorded SQL.
The agent uses these failures from both new and historical witnesses to modify the generator and its assignments, then repeats validation until all required checks pass or the evolution budget is exhausted.

\section{System Implementation}
\label{sec:implementation}
We implement \SystemName on \SQLancerPP as a multi-agent system with a control loop.
While, in principle, a human could provide the target statements, we aimed to fully automate the process.

\begin{figure}[t]
  \centering
  \includegraphics[width=0.45\textwidth]{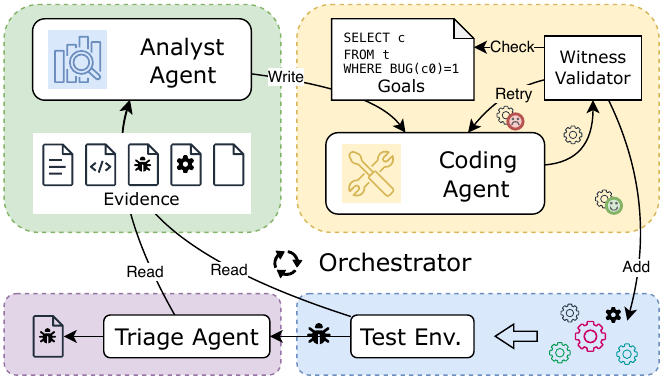}
  \caption{System overview of \SystemName. The analyst agent derives capability goals from DBMS evidence. The coding agent modifies the generator, and the validator checks the witnesses. The orchestrator continues the control loop. DBMS testing supplies evidence for subsequent analysis.}
  \Description{A diagram of a feedback loop with three regions. In the analysis region, an analyst agent reads evidence such as tests, code, bug reports, and execution artifacts and writes capability goals. In the coding region, goals enter a coding agent and the validator. The validator returns failed checks for retry and adds validated source modifications to the test environment. The test environment contains several DBMSs and testing components and feeds new evidence back to the analyst agent.}
  \label{fig:overview}
\end{figure}

\paragraph{Agent workflow.}
Figure~\ref{fig:overview} shows how \SystemName coordinates the analyst and coding agents with the evolved testing tool.
The analyst agent specifies capability goals containing target queries, and the coding agent modifies the generator and constructs generation witnesses to complete the goals.
This separation is motivated by studies of LLM-based testing showing that tests generated from incorrect code can repeat the implementation's incorrect assumptions~\cite{konstantinou2026codingbeforetesting,zhao2026misguidance}.
This separation also allows the agents to work in parallel. For example, while the analyst examines the target DBMS's source code, the coding agent can extend the generator using previously prepared goals.
The validator independently replays these witnesses against target queries and returns failures as feedback to the coding agent.
The orchestrator determines goal completion from the validation results and directs further testing with the accepted generator.
Testing results guide subsequent analysis, forming a feedback loop.

\subsection{Analysis}
\label{sec:goal-constraints}
The analyst agent examines DBMS source code, regression tests, bug reports, or prior testing results and specifies capability goals from these sources.
These goals guide the coding agent toward SQL features that the generator should support.

\paragraph{Capability goals.}
A \emph{capability goal} describes a SQL feature to be added using different variants of concrete target statements containing that feature.
Specifically, each goal contains a database setup (\emph{e.g.}, statements to create the tables) and distinct variants of target queries that witness replay should match.
For the row value \texttt{IN} feature, variants can vary the aggregate from \texttt{MIN} to \texttt{MAX} and change the columns in the row value.
These target statements should satisfy the constraints of the generator under evolution.
For example, nondeterministic features integrated into a TLP predicate may lead to false alarms; thus, the analyst should avoid incorporating such features.
We discuss this in more detail in Section~\ref{sec:discussion}.

\paragraph{Target constraints.}
The constraints guide target synthesis before the coding agent modifies the generator.
We provide the analyst agent with a natural language description of the supported target forms and the existing test oracle's requirements.
Listing~\ref{lst:oracle-prompt} pairs a simplified TLP oracle with the corresponding instructions.
First, a target should fit the oracle's supported query form.
For the WHERE variant of TLP, the analyst supplies a base query and a Boolean predicate from which the oracle constructs the partitions; an arbitrary SQL statement need not fit this form.
New features can appear in the predicate or the shared \texttt{FROM}/\texttt{JOIN} structure.
Second, the target should execute successfully on the selected DBMS with the supplied schema and data, so that the coding agent receives a valid example of the requested feature.
Third, the analyst should avoid features that can invalidate the target oracle.
This is a best-effort constraint---before adopting a target as a goal, the system independently executes its test case and checks that the base result equals the combined partition results.

\begin{listing}
  \caption{A simplified TLP oracle (top) and illustrative analyst agent constraints (bottom).}
  \label{lst:oracle-prompt}
  \begin{lstlisting}[style=generatorJava,numbers=none,basicstyle=\ttfamily\footnotesize,xleftmargin=1em,escapeinside={(*@}{@*)}]
Query q = select(columns, tables); (*@\colorbox{blue!12}{\strut [1]}@*)
Expr p = deterministicPredicate(); (*@\colorbox{green!12}{\strut [2]}@*)
Rows parts = concat(run(q, p), run(q, not(p)), run(q, isNull(p)));
assertSameMultiset(run(q), parts); (*@\colorbox{orange!18}{\strut [3]}@*)
  \end{lstlisting}
  \centering\smallskip
  {\small $\downarrow$\quad Constraints described to the analyst agent}\par
  \smallskip
  \begin{minipage}{\linewidth}
    \small
    \colorbox{blue!12}{\strut [1]} Provide the schema and data; use the same base query.\par
    \colorbox{green!12}{\strut [2]} Keep the predicate deterministic across executions.\par
    \colorbox{orange!18}{\strut [3]} Check that the combined partitions equal the base result.
  \end{minipage}
\end{listing}

\subsection{Coding}
The coding agent modifies \SQLancerPP and constructs witnesses for the targets of one goal, as illustrated in Section~\ref{sec:approach}.
Each modification attempt starts from the last accepted generator and runs in an isolated environment.
The witness mechanism validates the modification by replaying the provided witnesses and checking them against the targets specified by the goal.
Any mismatch or exception is used as feedback to help the coding agent modify the generator and witnesses.
The current implementation runs replay through the existing oracle execution path and also requires a passing oracle outcome, in addition to the target comparison.

\subsection{Controller}
\label{sec:pipelined-testing}
The controller pipelines analysis, coding, and testing.
The analyst agent prepares further goals while the coding agent modifies the generator and a background fuzzer tests accepted generators.

\paragraph{Testing schedule.}
Separating the analyst agent from the coding agent allows goal preparation to proceed alongside implementation.
Witness replay checks the requested targets directly, allowing validation to proceed without waiting for a random testing segment to finish.
After replay and regression checks pass, the orchestrator accepts the modified generator and queues it for testing.
The fuzzer switches between queued generators at segment boundaries and runs without interaction with LLMs.
A controller agent uses code coverage, execution statistics (\emph{e.g.}, query throughput), and completed goals to select the next goal and guide further analysis.

\paragraph{Handling frequent bugs.}
Frequent bug-inducing test cases can saturate testing without exposing further bugs, which is a common issue in automated testing~\cite{chen2013taming, donaldson2021testcase}.
A single SQL feature may repeatedly trigger known bugs, consuming testing time that could otherwise explore other features.
After each fuzzing segment, a triage agent analyzes the reported bugs and identifies whether changes to the generator are needed.
Large numbers of bug-inducing test cases increase triage effort; a repair agent thus modifies the generator to avoid recurring triggers.

\begin{figure}[t]
  \centering
  \includegraphics[width=\linewidth]{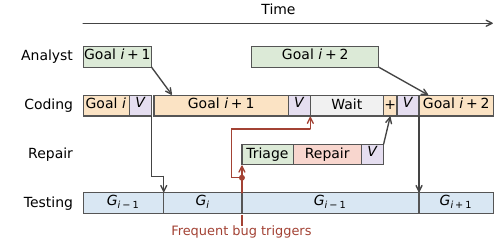}
  \caption{Pipelined evolution. $V$ denotes regression checks, $+$ denotes the code merge, and $G_i$ denotes the $i$-th validated generator.}
  \Description{Four compact lanes show analysis, coding, repair, and testing. Analysis prepares goal i plus one, then later goal i plus two. Testing switches from G i minus one to G i. At the end of the G i segment, one frequent-trigger event initiates triage and causes the coding result to wait after its ongoing work completes. Testing returns to G i minus one during repair. The repair is validated and mechanically merged with the coding result; the combined generator is validated before testing switches directly from G i minus one to G i plus one. Coding continues with goal i plus two, which is still in progress.}
  \label{fig:pipeline}
\end{figure}

Figure~\ref{fig:pipeline} illustrates how repair fits into the pipeline.
When bug-inducing test cases exceed a certain threshold (\emph{e.g.}, five for a testing segment), the orchestrator switches testing to a previously tested generator that did not show this behavior.
The coding agent continues its current modification, but the completed result waits for repair before integration.
Meanwhile, the repair agent modifies the generator that triggers the bugs, for example, by temporarily disabling the generation of certain SQL features that trigger known bugs.
The orchestrator then mechanically merges the repair with the coding agent's changes.
Any conflicts are resolved by the coding agent.
The combined generator is rebuilt and passes regression checks before entering another fuzzing segment.
Testing continues with the earlier generator until the combined generator is ready; if no suitable earlier generator is available, it waits throughout repair.
Section~\ref{sec:practicality} examines how these activities overlap in a complete run.

\section{Evaluation}
\label{sec:evaluation}

We evaluated \SystemName in terms of bug-finding effectiveness, feature acquisition, performance compared with baselines, and practicality.
First, we evaluate the bug-finding effectiveness of \SystemName on three popular DBMSs.
Second, we measure whether witness replay helps coding agents acquire and preserve requested features.
Third, we compare the evolved generator with state-of-the-art DBMS testing tools.
Finally, we study how \SystemName evolves \SQLancerPP and tests DBMSs from a bug-finding request.

\paragraph{Baselines.}

\begin{sloppypar}
We compare \SystemName with ShQveL, SQLancer, and the initial \SQLancerPP generator.
We also include the DBMSs' native regression tests as a reference.
ShQveL is the closest baseline because it also uses LLMs to extend test case generation, but learns SQL fragments rather than modifying generator code~\cite{zhong2025shqvel}.
SQLancer provides manually implemented generators and a platform for test oracles such as TLP, DQP, and CODDTest~\cite{rigger2020finding,ba2024keep,zhang2025coddtest}.
\SQLancerPP provides a general generator across SQL dialects and underlies ShQveL and JIR~\cite{zhong2025scaling,zhong2025shqvel,xiang2026jir} and serves as the initial generator.
Test suites can supply seeds for mutation-based tools and already exceed SQL generators in coverage~\cite{zhong2025scaling,zhong2025shqvel}.
\end{sloppypar}

\paragraph{Setup.}
\begin{sloppypar}
We conducted the experiments using a server with a \ExperimentCPU{} and \ExperimentMemory{} of memory running \ExperimentOS{}.
For feature acquisition, each measurement used \FeatureFuzzThreads{} generator threads, a CPU quota of \FeatureMeasurementCPUs{} cores, and a \FeatureMeasurementMemory{} memory limit for \FeatureFuzzSeconds{} seconds per seed.
For baseline comparison, each tool received \BaselineRunBudget{}, a CPU quota of \BaselineCPUs{} cores, and \BaselineMemory{} of memory during DBMS testing.
Each generator used \BaselineGeneratorThreads{} thread; coverage replay for SQLite and DuckDB used \BaselineReplayWorkers{} workers within the same allocation.
The practicality case study used \DuckdbPracticalityFuzzThreads{} fuzzer threads and a \DuckdbPracticalityBudgetHours{}-hour budget for admitting new work.

\end{sloppypar}

\paragraph{DBMS selection.}
We select three open-source DBMSs with different workload focuses.
SQLite and PostgreSQL primarily target transactional workloads, while DuckDB primarily targets analytical workloads.
SQLite and DuckDB run as embedded systems, whereas PostgreSQL runs as a server.
We use SQLite, DuckDB, and PostgreSQL for bug finding and baseline comparison.
For the coverage comparison, we selected SQLite \BaselineSQLiteVersion{}, DuckDB \BaselineDuckDBVersion{}, and PostgreSQL \BaselinePostgreSQLVersion{}.
SQLite and PostgreSQL are among the \emph{most challenging} targets for finding logic bugs.
SQLite's TH3 test suite achieves 100\% branch coverage of its core library.\footnote{\url{https://www.sqlite.org/testing.html\#test_coverage}}
TLP implemented in SQLancer found no logic bugs in PostgreSQL~\cite{rigger2020finding}.
Among the DBMS testing papers on logic bug detection published within the last five years at SIGMOD and VLDB, only four report confirmed logic bugs in SQLite~\cite{zhang2025coddtest,zhang2026epsc,song2024radar,xiang2026jir} and four in PostgreSQL~\cite{deng2025edc,lai2025srs,mang2026argus,zhang2026epsc}.
We excluded MySQL because earlier testing campaigns reported that many known bugs remained unfixed~\cite{zhong2026dbugscribe}, making it difficult to assess the contribution of new testing methods.
We use DuckDB as a case study to evaluate the practicality of \SystemName.

\subsection{Bug-Finding Effectiveness}
\label{sec:bug-finding}

\begin{table}[t]
  \centering
  \setlength{\tabcolsep}{4pt}
  \caption{\SystemName allowed us to find and report \CampaignBugs{} confirmed bugs in three DBMSs, of which \CampaignFixed{} have been fixed. The bugs include \CampaignLogic{} logic bugs and \CampaignOther{} other defects. The table also shows the DBMS size in KLOC.}
  \label{tab:bug-campaign}
  \begin{tabular}{@{}lrrrrr@{}}
    \toprule
    DBMS & LOC (K) & Confirmed & Fixed & Logic & Other \\
    \midrule
    SQLite & \SQLiteLOC & \SQLiteBugs & \SQLiteFixed & \SQLiteLogic & \SQLiteOther \\
    DuckDB & \DuckDBLOC & \DuckDBBugs & \DuckDBFixed & \DuckDBLogic & \DuckDBOther \\
    PostgreSQL & \PostgresLOC & \PostgresBugs & \PostgresFixed & \PostgresLogic & \PostgresOther \\
    \midrule
    \textbf{Total} & & \CampaignBugs & \CampaignFixed & \CampaignLogic & \CampaignOther \\
    \bottomrule
  \end{tabular}
\end{table}

We evaluated \SystemName in an unstructured testing campaign on SQLite, DuckDB, and PostgreSQL, following the practice of prior DBMS testing work~\cite{rigger2020finding,zhong2025scaling}.
We used TLP to detect incorrect query results and monitored failures such as crashes and internal errors.
The system reduced and prioritized the bug-inducing test cases, which we further examined before reporting them to developers.
We checked earlier issue reports and fixes for duplicates and excluded reports that developers classified as intended behavior from the bug count.
The results include bugs found after generator modifications, through configuration changes, and through direct analysis of DBMS source code by an agent.

\paragraph{Bug statistics.}
Table~\ref{tab:bug-campaign} summarizes the bug statistics for each DBMS.
In total, we submitted \CampaignReports{} reports, of which \CampaignBugs{} describe bugs confirmed by developers and \CampaignIntended{} describes intended behavior.
Among the confirmed bugs, \CampaignFixed{} have been fixed.
\CampaignLogic{} bugs cause incorrect query results, while \CampaignOther{} cause unexpected errors or crashes.
All \SQLiteBugs{} confirmed SQLite bugs have been fixed.
The DuckDB developers confirmed \DuckDBBugs{} bugs and fixed \DuckDBFixed{}; the PostgreSQL developers confirmed \PostgresBugs{} and fixed \PostgresFixed{}.
The five DuckDB reports from the practicality case study concern nested types, casts, statistics propagation, and predicate optimization; they are included in the confirmed bug counts.
These results demonstrate that \SystemName can find bugs that DBMS developers consider worth fixing, including \SQLiteLogic{} logic bugs in SQLite and \PostgresLogic{} in PostgreSQL despite extensive prior testing.
For example, CODDTest reported six confirmed SQLite logic bugs~\cite{zhang2025coddtest}, while EDC, SRS, Argus, and EPSC reported one, two, one, and one confirmed PostgreSQL logic bugs, respectively~\cite{deng2025edc,lai2025srs,mang2026argus,zhang2026epsc}.
These findings show that extending the generator can expose new bugs in well-tested DBMSs even when using a single existing test oracle.

\paragraph{Bug feature analysis}
We further characterize how acquired features combine in these bug-inducing test cases.
Among the 44 bugs, 19 required at least one evolved feature, 8 required two, and 14 required at least three.
For each evolution run in our bug-finding campaign, the first bug appeared at evolution iteration 23.7 on average.
The seed generator could find three DuckDB bugs without modifications.
These counts reflect the features in each bug-inducing test case after best-effort reduction.
Fewer features might still trigger the bug, but the same oracle might not detect it.
We discuss three logic bugs and the generator capabilities needed to expose them.

\begin{lstlisting}[style=bugSQL,
  caption={Adding a \texttt{WHERE} predicate changes a SQLite query's result from no rows to three rows.},
  label={lst:sqlite-json-each}]
CREATE TABLE t0(c0 TEXT);
INSERT INTO t0 VALUES ('[1,2,3]');

SELECT value FROM json_each(t0.c0),
  t0 RIGHT JOIN (SELECT 1) sub0 ON true;
-- Returns no rows

SELECT value FROM json_each(t0.c0),
  t0 RIGHT JOIN (SELECT 1) sub0 ON true
WHERE t0.c0 IS NOT NULL;
-- Returns 1, 2, 3
\end{lstlisting}
\paragraph{SQLite bug involving \texttt{json\_each()} and \texttt{RIGHT JOIN}.}
Listing~\ref{lst:sqlite-json-each} shows a SQLite bug where adding a \texttt{WHERE} predicate changes a query's result from no rows to three rows.
The initial generator could not use JSON table-valued functions in the base query's \texttt{FROM} clause.
The coding agent added these functions, exposed their output columns to predicate generation, and extended data generation with valid JSON values.
The following testing phase then combined the new capability with existing \texttt{RIGHT JOIN} generation and exposed the inconsistency.

\begin{lstlisting}[style=bugSQL,
  caption={An indexed predicate returns one row from an empty DuckDB sample.},
  label={lst:duckdb-tablesample-index}]
CREATE TABLE t0(c1 BOOLEAN);
CREATE INDEX i0 ON t0(c1);
INSERT INTO t0 VALUES (NULL), (true);

SELECT count(*) FROM t0 TABLESAMPLE SYSTEM (66%) REPEATABLE (209);
-- 0
SELECT count(*) FROM t0 TABLESAMPLE SYSTEM (66%) REPEATABLE (209)
WHERE c1 IN (SELECT c1 FROM t0 GROUP BY ALL);
-- Expected: 0; Actual: 1
\end{lstlisting}

\paragraph{DuckDB bug in indexed sampled scans.}
\begin{sloppypar}
Listing~\ref{lst:duckdb-tablesample-index} shows a DuckDB wrong result: the sampled table contains no rows, but filtering it with an \texttt{IN} subquery returns one.
The coding agent added \texttt{TABLESAMPLE} to the generator's \texttt{FROM} clauses, and the subsequent testing phase combined it with an indexed Boolean column and an existing \texttt{IN}-subquery predicate.
Without the index, the filtered query also returns zero rows.
This case shows a newly acquired generator capability combining with existing schema and predicate generation to expose a DuckDB wrong result during subsequent testing.
\end{sloppypar}

The bug-inducing test case combining different features was generated after several iterations of generator evolution.
The coding agent added \texttt{IN}-subquery predicates in iteration~5, extended these subqueries with \texttt{GROUP BY ALL} in iteration~18, and added \texttt{TABLESAMPLE} in iteration~30.
Testing in iteration~32 combined these capabilities over an indexed Boolean column and exposed the result mismatch.
The index came from existing schema generation.

\begin{lstlisting}[style=bugSQL,
  caption={PostgreSQL counts two rows although the same \texttt{IN} predicate matches only one.},
  label={lst:postgres-collation}]
CREATE COLLATION ci (provider = icu,
  locale = 'und-u-ks-level1', deterministic = false);
CREATE TABLE t_lhs(c1 text COLLATE ci);
CREATE TABLE t_rhs(c0 text);
INSERT INTO t_lhs VALUES ('a'), ('x'), ('y');
INSERT INTO t_rhs VALUES ('a'), ('a');
ANALYZE t_lhs;
ANALYZE t_rhs;
INSERT INTO t_rhs VALUES ('A');

SELECT c1, c1 IN (SELECT c0 FROM t_rhs) AS p FROM t_lhs;
-- a | t
-- x | f
-- y | f
SELECT count(*) FROM t_lhs WHERE c1 IN (SELECT c0 FROM t_rhs);
-- Expected: 1, Actual: 2
\end{lstlisting}

\paragraph{PostgreSQL bug in collation handling.}
Listing~\ref{lst:postgres-collation} shows a PostgreSQL wrong result: the \texttt{IN} predicate matches only \texttt{\textquotesingle a\textquotesingle} in the projection, but the filtered query counts two rows.
The left column's nondeterministic ICU collation equates \texttt{\textquotesingle a\textquotesingle} and \texttt{\textquotesingle A\textquotesingle}, while the right column uses the default collation.
The planner deduplicated the semijoin's right side under the default rather than the join collation, retaining both values and emitting the match twice; the fix uses the join collation.
The initial generator could not create such collations.
The coding agent added collation creation and assignment to text columns, allowing existing \texttt{IN}-subquery generation to expose the defect.

\paragraph{Bug reachability analysis.}
These three cases illustrate why reaching new bugs requires extending query and schema generation beyond learning SQL fragments or discovering test oracles.
First, the SQLite case requires a JSON table-valued function in the \texttt{FROM} clause and its output columns in subsequent predicates; ShQveL's expression sketches cannot supply this integration, and Argus does not support SQLite~\cite{zhong2025shqvel,mang2026argus}.
Second, the DuckDB case combines \texttt{TABLESAMPLE}, an index, and an \texttt{IN} subquery, which is beyond ShQveL's expression sketches.
Third, the PostgreSQL case requires creating a nondeterministic ICU collation and assigning it to a column before generating the \texttt{IN} predicate; learning expressions alone does not add this schema support.
In principle, Argus's table placeholders can represent the required table expressions with suitable snippets and could expose the latter two bugs; however, this requires generating corresponding snippet pools.
Reaching features such as custom collations still requires an efficient schema generator.
\SystemName instead integrates LLM knowledge into reusable query and schema generation rules, allowing subsequent random testing to combine the acquired features across newly generated database states without further LLM calls for SQL generation.

\subsection{Feature Acquisition}
\label{sec:feature-acquisition}
We evaluated whether witness replay helps coding agents implement requested SQL features in the generator.
We compared evolution with and without witnesses in two session settings and measured task completion and query variety.

\begin{figure}[t]
  \centering
  \includegraphics[width=\columnwidth]{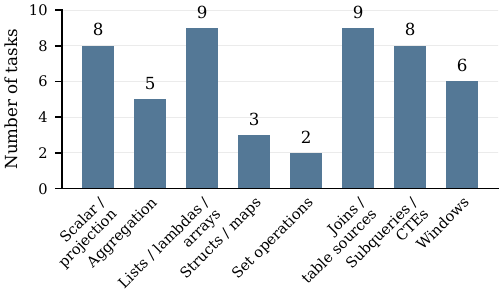}
  \caption{The \BenchmarkTasks{} query targets grouped by the generator functionality they primarily test. Bar labels give target counts.}
  \Description{Eight bars show 8 scalar and projection targets, 5 aggregation targets, 9 list, lambda and array targets, 3 struct and map targets, 2 set-operation targets, 9 join and table-source targets, 8 subquery and CTE targets, and 6 window targets.}
  \label{fig:feature-benchmark}
\end{figure}

\paragraph{Tasks.}
We selected \BenchmarkTasks{} query targets from distinct extension goals mined from DuckDB's documentation and regression tests after inspecting the generator's source code for missing features.
Each task supplied a target query, a schema, and data, which we checked against DuckDB.
The targets cover eight categories, including expressions, aggregation, nested values, table sources, subqueries, and windows (see Figure~\ref{fig:feature-benchmark}).
They require changes to generation rules, type handling, and scope handling.
Listing~\ref{lst:feature-task} shows a condensed prompt for list subscripting.
The initial generator supported function-based list access but lacked postfix subscripts.
We instructed the agent to integrate typed subscripts into random generation over live schemas.
Both conditions received the same target and schema settings, together with instructions to vary columns and constants, preserve earlier capabilities, and sample comparable generation choices uniformly.

\begin{lstlisting}[style=featureTask,caption={A condensed feature-acquisition task for list subscripting.},label={lst:feature-task}]
Schema: t0(c0 INT[], c1 INT)
Target: SELECT * FROM t0 WHERE (((t0.c0)[1])>(t0.c1))
Request:
  Implement typed list subscripts in random generation.
  Do not hard-code the target SQL.
  Vary columns and constants; preserve earlier features.
  Sample comparable legal generation choices uniformly.
\end{lstlisting}

\paragraph{Settings.}
We used OpenCode with \texttt{deepseek-flash} for all settings, starting from the same generator functionality and using development guidance derived from the same skill.
We evaluated with and without witness under multi-session and one-session.
These settings let us compare witness feedback during incremental feature acquisition across conversations and development of all requested features in one conversation.
In multi-session evolution, the agent received one task per new conversation and started from the last accepted generator, with acceptance feedback and handoff notes.
Each task allowed one coding attempt; a rejected modification left the accepted generator unchanged.
In one-session evolution, the agent received all \BenchmarkTasks{} tasks in one conversation and managed development and testing, including witness validation in the with-witness condition.

\paragraph{Metrics.}
We measured task completion over time and query variety in the final generator.
A task is complete when the generator reproduces its target query through the specified testing interface that the agent is required to implement.
This query matching ignored whitespace, unquoted word case, and a terminal semicolon.
We evaluated both session settings against all \BenchmarkTasks{} target queries.
At each multi-session checkpoint, task completion is the fraction of all targets reproduced by the current generator, including targets introduced in earlier tasks.
For one-session evolution, we applied this metric to the final generator after the agent exited.
This metric captures both newly acquired features and their preservation through subsequent changes.

We also measured query variety through the distinct AST edges in random queries containing at least one target feature.
We examined each query's complete AST, including nested subqueries, and deduplicated edges by their signatures.
Each signature consists of the parent node label, child role, and child node label; it preserves operator and function identities but ignores identifiers and literal values.
We sampled 10,000 queries in total from each final generator across \FeatureMeasurementSeeds{} fixed random seeds.
For each seed, we divided the number of distinct edge signatures in queries containing a target feature by the number of all generated, parseable statements, then averaged these ratios across seeds.

\paragraph{Results.}
Table~\ref{tab:feature-final} shows that witnesses improved task completion and query variety in multi-session evolution.
The final generator reproduced 47 of 50 targets (\MultiWithSuccessrate{}) with witnesses, compared with 37 (\MultiWithoutSuccessrate{}) without witnesses.
Witnesses helped complete targets involving ordered-set aggregation, recursive CTEs, and window exclusion frames.
Query variety reached \MultiWithASTedges{} distinct AST edges per \texttt{SELECT} with witnesses, compared with \MultiWithoutASTedges{} without witnesses.
In one-session evolution, the with-witness agent reproduced all 50 targets (\AutonomousWithSuccessrate{}), but query variety was only \AutonomousWithASTedges{}.
We inspected the source code and found that it hardcoded the target queries for witness replay, yielding little variety involving the requested features during random generation.
Without witnesses, the agent reproduced two targets (\AutonomousWithoutSuccessrate{}), while query variety reached \AutonomousWithoutASTedges{}.
Without witnesses, multi-session evolution achieved less query variety than one-session evolution because it also include hardcode \texttt{SELECT}s for some tasks (\emph{e.g.}, \texttt{list\_aggregate}), while one-session evolution integrated features into recursive expressions and \texttt{FROM} components, enabling more varied combinations.

\begin{figure}[t]
  \centering
  \includegraphics[width=\columnwidth]{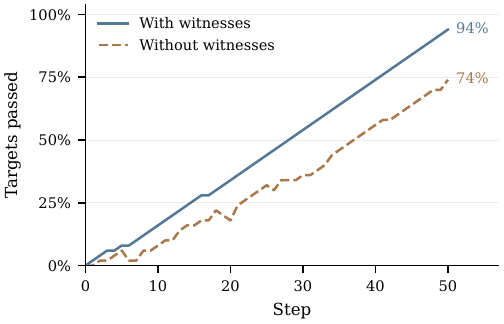}
  \caption{Task completion during multi-session
evolution with and without witnesses. Each checkpoint shows the fraction of all \BenchmarkTasks{} targets reproduced by the current generator.}
  \Description{Two curves show target pass rates over 50 tasks by checkpoint. With witnesses, the final pass rate is 94 percent. Without witnesses, the pass rate declines at several checkpoints and ends at 74 percent.}
  \label{fig:feature-acquisition}
\end{figure}

Figure~\ref{fig:feature-acquisition} shows how multi-session evolution acquired and retained generator capabilities.
The with-witness condition preserved all previously passing registered tests, while the without-witness curve declined at several checkpoints as earlier tests failed.
The with-witness condition did not achieve 100\% task completion because the agent failed to complete three tasks within the single allowed attempt. In the full evolution process, witness feedback identifies test failures and guides further attempts.

\begin{table}[t]
  \centering
  \setlength{\tabcolsep}{4pt}
  \caption{Final task completion and query variety with OpenCode and \texttt{deepseek-flash}.}
  \label{tab:feature-final}
  \begin{tabular}{@{}llrr@{}}
    \toprule
    Session & Witness & Completion & Variety \\
    \midrule
    Multi & Yes & \MultiWithSuccessrate & \MultiWithASTedges \\
    Multi & No & \MultiWithoutSuccessrate & \MultiWithoutASTedges \\
    One & Yes & \AutonomousWithSuccessrate & \AutonomousWithASTedges \\
    One & No & \AutonomousWithoutSuccessrate & \AutonomousWithoutASTedges \\
    \bottomrule
  \end{tabular}
\end{table}

\subsection{Baseline Comparison}
\label{sec:baseline-comparison}
We compared the line and branch coverage of the evolved generator with ShQveL, \SQLancerPP, \SQLancer, and the public native test suites on SQLite,\footnote{The TH3 test suite that achieves 100\% branch coverage is not publicly available.} DuckDB, and PostgreSQL.
Our main comparison is with ShQveL, which extends \SQLancerPP by learning SQL fragments rather than modifying generator code.
The comparison with the initial \SQLancerPP generator measures the effect of generator evolution under the same test oracle.
Coverage provides an indirect indication of feature coverage, but is only a loose proxy for logic bug-finding effectiveness~\cite{zhong2025scaling}.

\paragraph{Methodology.}
We compared fixed generators to measure their DBMS coverage after learning, without further LLM calls during testing.
For \SystemName, we evolved each generator for 24 hours using OpenCode with \texttt{deepseek-flash}; for ShQveL, we allocated the same preparation budget to learning SQL fragments.
We then fixed the evolved generators and learned fragments before testing each tool on the same instrumented DBMS builds under the resource limits described in Setup.
No further LLM calls were made during testing.
The comparison measures the testing capability kept after preparation, rather than coverage accumulated while the generators were changing.
Preparation differed across tools; the subsequent DBMS-testing budget was matched.
We ran regression tests of each DBMS once as a coverage reference.

\begin{table}[t]
  \centering
  \setlength{\tabcolsep}{3pt}
  \caption{Line and branch coverage on SQLite, PostgreSQL,
and DuckDB after 24 hours, comparing the evolved \SystemName generator against \SQLancerPP{},
ShQveL, SQLancer, and native test suites.}
  \label{tab:baselines}
  \begin{tabular}{@{}lrrrrrr@{}}
    \toprule
     & \multicolumn{2}{c}{SQLite} & \multicolumn{2}{c}{DuckDB} & \multicolumn{2}{c}{PostgreSQL} \\
    \cmidrule(lr){2-3}\cmidrule(lr){4-5}\cmidrule(lr){6-7}
    Approach & Line & Branch & Line & Branch & Line & Branch \\
    \midrule
    \SystemName & \BaselineSQLiteGungnirLine & \BaselineSQLiteGungnirBranch & \BaselineDuckDBGungnirLine & \BaselineDuckDBGungnirBranch & \BaselinePostgreSQLGungnirLine & \BaselinePostgreSQLGungnirBranch \\
    \SQLancerPP{}& \BaselineSQLiteSqlancerPPLine & \BaselineSQLiteSqlancerPPBranch & \BaselineDuckDBSqlancerPPLine & \BaselineDuckDBSqlancerPPBranch & \BaselinePostgreSQLSqlancerPPLine & \BaselinePostgreSQLSqlancerPPBranch \\
    ShQveL & \BaselineSQLiteShqvelLine & \BaselineSQLiteShqvelBranch & \BaselineDuckDBShqvelLine & \BaselineDuckDBShqvelBranch & \BaselinePostgreSQLShqvelLine & \BaselinePostgreSQLShqvelBranch \\
    \SQLancer & \BaselineSQLiteSqlancerLine & \BaselineSQLiteSqlancerBranch & \BaselineDuckDBSqlancerLine & \BaselineDuckDBSqlancerBranch & \BaselinePostgreSQLSqlancerLine & \BaselinePostgreSQLSqlancerBranch \\
    \midrule
    Test suite & \TestSuiteSQLiteLine & \TestSuiteSQLiteBranch & \TestSuiteDuckDBLine & \TestSuiteDuckDBBranch & \TestSuitePostgreSQLLine & \TestSuitePostgreSQLBranch \\
    \bottomrule
  \end{tabular}
\end{table}

\paragraph{Results.}
\begin{sloppypar}
Table~\ref{tab:baselines} reports line and branch coverage.
On SQLite, DuckDB, and PostgreSQL, \SystemName achieved \BaselineSQLiteGungnirBranch{}\%, \BaselineDuckDBGungnirBranch{}\%, and \BaselinePostgreSQLGungnirBranch{}\% branch coverage, respectively.
ShQveL covered \BaselineSQLiteShqvelBranch{}\%, \BaselineDuckDBShqvelBranch{}\%, and \BaselinePostgreSQLShqvelBranch{}\% on the same systems.
Compared with the initial \SQLancerPP generator, \SystemName increased branch coverage by \BaselineSQLiteBranchGain{}, \BaselineDuckDBBranchGain{}, and \BaselinePostgreSQLBranchGain{} percentage points, respectively.
It also covered more lines and branches than the measured generator baselines on all three DBMSs.
The test suites achieved \TestSuiteDuckDBBranch{}\% and \TestSuitePostgreSQLBranch{}\% branch coverage on DuckDB and PostgreSQL, respectively, exceeding all evaluated generators, including \SystemName.
The gap is expected because evaluated generators focus on logic bug detection, and the query forms are constrained by the test oracle, while the test suites cover a broader range of operations, including schema changes and administrative commands.
\SystemName achieved higher coverage on SQLite than on DuckDB and PostgreSQL; one possible explanation is that SQLite's smaller codebase makes it easier for the analyst to identify SQL features from source code. A smaller set of SQL features could also make broader coverage easier to achieve.
\end{sloppypar}

\begin{figure}[t]
  \centering
  \includegraphics[width=\linewidth]{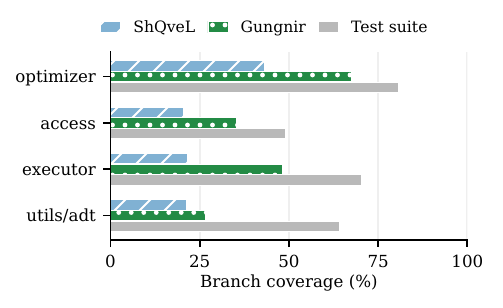}
  \caption{PostgreSQL branch coverage under \texttt{src/backend}. We select the four top-level directories with the largest gains over ShQveL. Each generator ran for 24 hours; the native regression suite ran once on the same build.}
  \Description{Grouped horizontal bars compare ShQveL, Gungnir, and the test suite in optimizer, access, executor, and utils/adt. The test suite has the highest coverage in each directory, followed by Gungnir and ShQveL.}
  \label{fig:coverage-components}
\end{figure}

\paragraph{Fine-grained coverage analysis.}
Figure~\ref{fig:coverage-components} compares PostgreSQL branch coverage for ShQveL, \SystemName, and the native test suite in selected source directories, with \texttt{utils} narrowed to its main component, \texttt{utils/adt}.
\SystemName covered 4,743 additional branches in \texttt{optimizer}, 4,108 in \texttt{access}, 3,500 in \texttt{executor}, and 2,503 in \texttt{utils/adt}, accounting for 63.0\% of its net increase over ShQveL.
These gains span query planning, data access, execution, and utility functions, although the test suite covers more branches in each directory.
The \texttt{utils/adt} directory contains 74.0\% of the branches in \texttt{utils} and accounts for 83.3\% of the test-suite gap in \texttt{utils}; it includes data-type functions and utilities such as query deparsing, for which the test suite covered 29,875 branches compared with 12,420 for \SystemName.
The suite also includes operations such as plan inspection and function management, illustrating the broader scope of native regression testing.
Thus, the remaining coverage gap reflects both operations absent from the generated workload and incomplete branch coverage within components that the generator already reaches.

\subsection{Practicality}
\label{sec:practicality}
We further evaluated the practicality of \SystemName through a case study on DuckDB.
We recorded how the system evolved the generator, tested the DBMS, and generated bug reports.
We used DuckDB nightly version with \DuckdbPracticalityLLM{} through OpenCode and issued the following request to the orchestrator agent: \emph{``Do not stop until you find \DuckdbPracticalityBugTarget{} unique, previously unknown bugs.''}
We counted the bug reports generated by final triage and measured elapsed time, recorded token use, and LLM cost through the end of that session.

\begin{figure}[t]
  \centering
  \includegraphics[width=\linewidth]{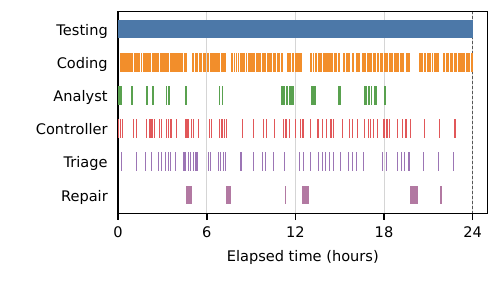}
  \caption{Evolution timeline for DuckDB. Each bar denotes an agent session or fuzzer segment; aligned bars show concurrent activity. Time starts at the first recorded activity, and the dashed line marks the end of testing. Final triage is omitted.}
  \Description{A Gantt chart with rows for the fuzzer, coding agent, analyst agent, controller agent, triage agent, and repair agent. Testing spans 23.99 hours and overlaps with agent sessions. Final triage is not shown.}
  \label{fig:duckdb-practicality}
\end{figure}

\paragraph{Execution.}
Figure~\ref{fig:duckdb-practicality} shows how generator evolution and DBMS testing overlapped in the DuckDB run.
The analyst agent examined DBMS source code and coverage results to identify additional SQL features, while the coding agent implemented the resulting goals.
The controller agent selected goals and adjusted testing as new evidence became available, and the triage agent examined failures alongside continued testing.
By the end of the run, the generator satisfied \DuckdbPracticalityAcceptedGoals{} SQL feature goals.
The run accepted \DuckdbPracticalityAcceptedModifications{} source commits, including repeated work after rejection; each passed the existing regression checks, including witness replay.
Over \DuckdbPracticalityTestingHours{} hours, the fuzzer tested the seed and \DuckdbPracticalityTestedRevisions{} evolved bundles in \DuckdbPracticalitySegments{} segments, executing \DuckdbPracticalityStatementsMillion{} million SQL statements.
The fuzzer was active for 23.92 hours within this run, while coding sessions occupied 17.39 hours.
These overlapping activities illustrate the pipeline in Section~\ref{sec:pipelined-testing}: testing an accepted generator need not wait for the coding session to finish.

\paragraph{Results.}
The DuckDB run reproduced nine kinds of bugs, from which the triage agent excluded four upstream duplicates and generated five bug reports.
We submitted these reports directly to DuckDB developers, who confirmed all of them.
This result also illustrates the importance of triage agents in filtering duplicates and upstream issues.
The whole run took \DuckdbPracticalityElapsedHours{} hours, including an extra final triage for generating reports and reproductions.
In total, the run used \DuckdbPracticalityTotalTokensMillion{} million tokens, at an LLM cost of \$\DuckdbPracticalityCostUSD{}.
Cached input accounted for 98.57\% of input tokens, so the total token count alone does not explain the recorded cost.
A longer evolution budget, such as one week, could allow the system to implement more SQL features and test their combinations for longer.

\section{Discussion and Limitations}
\label{sec:discussion}

\paragraph{Scope of witness validation.}
Witness replay checks whether the generator can reproduce specific target statements.
However, it has three limitations.
First, it does not guarantee how frequently the feature is generated during random testing.
For example, a generator may be able to produce a target statement but need choices that are rarely selected during random generation.
Second, it does not guarantee that the new feature can be combined correctly with other features.
For example, a generator may be able to generate a new function requiring an argument of a specific data type, but such constraints may not be encoded in the generator's random generation logic.
Third, it requires that the target statement is valid and satisfies the test oracle's input requirements; witness replay cannot check whether the target statement is semantically correct.

\paragraph{False alarms.}
The evolved generator can still generate queries that cause false alarms, that is, cases in which the test oracle reports a bug although the DBMS behaves correctly.
For example, SQLite permits ambiguous aggregate queries whose bare columns take values from an arbitrarily selected row within a group.
Using these queries in the TLP base query can yield different results, which is expected behavior.
These cases were uncommon in our testing, and all were identified and removed by the triage agent.
The analyst can record these false positives and avoid similar target queries in later iterations.

\paragraph{Amortizing evolution cost.}
The LLM cost of generator evolution can be spread over repeated DBMS testing.
An agent that directly searches for bugs uses LLM calls to generate and refine test cases; \SystemName instead uses these calls to extend a reusable generator.
Once evolution ends, the generator can continue random testing without further LLM calls.
In the DuckDB case study (Section~\ref{sec:practicality}), the approximately \DuckdbPracticalityBudgetHours{}-hour run completed \DuckdbPracticalityAcceptedGoals{} SQL feature goals at a recorded LLM cost of \$\DuckdbPracticalityCostUSD{}, including analysis, coding, repair, and triage.
In the coverage experiment (Section~\ref{sec:baseline-comparison}), we tested the fixed DuckDB generator without further LLM calls and measured \BaselineDuckDBBranchGain{} percentage points more branch coverage than the initial \SQLancerPP generator.

\section{Related Work}
\label{sec:related}

\paragraph{DBMS testing.}
Automated DBMS testing combines test case generation with test oracles that detect unexpected behavior.
For test case generation, Squirrel~\cite{zhong2020squirrel} and SQLRight~\cite{liang2022detecting} mutate existing SQL statements and use validity and coverage feedback to guide testing.
QPG~\cite{ba2024qpg} mutates the database state to obtain query plans that have not been seen during testing.
\SQLancerPP~\cite{zhong2025scaling} uses a shared adaptive generator to generate SQL statements across dialects.
For test oracles, PQS~\cite{rigger2020testing} checks query results against known rows.
TLP~\cite{rigger2020finding} generates three queries that partition the rows of an original query and checks whether their combined results match its result.
NoREC~\cite{rigger2020detecting} compares the result of a query that can be optimized with that of an equivalent query designed to avoid those optimizations.
DQP~\cite{ba2024keep} executes a query using different query plans and compares their results.
Radar~\cite{song2024radar} removes database metadata to check for incorrect query optimizations.
JIR~\cite{xiang2026jir} uses the results of related join types to derive the expected result of a join.
LLMs have been used to improve DBMS testing.
ShQveL~\cite{zhong2025shqvel} adds validated SQL fragments to a generator that reuses them during testing.
QTRAN~\cite{lin2025qtran} uses LLMs to translate queries across SQL dialects and adapts mutations to preserve the relations required by metamorphic testing.
Argus~\cite{mang2026argus} prompts an LLM to generate abstract query pairs and checks their equivalence with a SQL equivalence prover before using them as test oracles.
Building on \SQLancerPP, \SystemName modifies the generator's source code to support new SQL features and uses existing test oracles to validate the generated statements.

\paragraph{LLMs for databases.}
LLMs have been used to improve query rewriting, configuration tuning, and workload generation.
LLM-R\textsuperscript{2}~\cite{li2024llmr2} uses an LLM to select rewrite rules from existing SQL rewriting systems.
QO-Verify~\cite{narasayya2026leveraging} uses a query optimizer to check the equivalence of LLM-generated rewrites.
SQLFlex~\cite{an2026dialect} parses SQL by integrating grammar-based parsing with LLM capabilities.
SQLStorm~\cite{schmidt2025sqlstorm} generates SQL workloads for benchmarking.
SQLBarber~\cite{lao2026sqlbarber} generates SQL workloads according to requirements supplied by the user.
Recently, LLM agents have been used to coordinate database tuning and SQL generation.
AgentTune~\cite{li2025agenttune} coordinates LLM agents that analyze workloads, select configuration parameters, and refine database configurations.
SysInsight~\cite{zhang2026sysinsight} combines static analysis with LLM reasoning to derive tuning rules from DBMS source code and refines these rules using performance feedback.
SQL-Factory~\cite{li2025sqlfactory} coordinates agents to generate SQL patterns, expand them into queries, and evaluate query quality for use in downstream tasks.
\SystemName uses coding agents to extend a DBMS test generator with new SQL features.

\paragraph{Agentic software evolution.}
LLMs can evolve software by modifying source code using execution feedback.
AlphaEvolve~\cite{novikov2025alphaevolve} uses LLMs to generate code changes and combines evolutionary search with automated evaluation to improve algorithms.
SkyDiscover~\cite{liu2026skydiscover} evolves programs using adaptive search strategies.
FrontierCS~\cite{mang2026frontiercs} is a benchmark that evaluates executable solutions to open-ended computer science problems.
IC3-Evolve~\cite{miao2026ic3evolve} modifies hardware model-checking heuristics and checks safety proofs and counterexample traces independently to validate the changes.
Such LLM-based software evolution has been applied to software testing tools.
ELFuzz~\cite{chen2025elfuzz} uses LLMs to modify generation-based fuzzers and coverage feedback to guide their evolution.
Gentoo~\cite{vikram2026gentoo} uses coding agents to implement and refine input generators, with optional predicate-level coverage feedback.
GenEvo~\cite{wei2026genevo} uses LLMs to solve constraints and adapt input generation logic to target program states.
\SystemName uses fixed SQL targets to validate generator modifications.
Witness replay directs the modified generator to generate these statements, and existing test oracles check their required behavior.
Replaying historical witnesses checks that subsequent modifications preserve previously witnessed target generation.

\paragraph{Controlled test generators.}
Prior work controls the choices made by a test generator to guide the inputs it generates.
Swarm testing~\cite{groce2012swarm} enables different subsets of features across tests.
Directed swarm testing~\cite{groce2016directed} uses feedback to guide feature selection.
Targeted property-based testing~\cite{loscher2017targeted} uses a fitness function supplied by the user to search for test inputs.
Zest~\cite{padhye2019zest} mutates the parameters supplied to a structured generator to guide testing.
Loaded Dice~\cite{tjoa2025tuning} adjusts the weights of generator choices to better match a specified distribution of generated values.
Parsing Randomness~\cite{goldstein2022parsing} treats a generator as a parser whose input is a sequence of random choices.
Reflective generators~\cite{goldstein2023reflecting} recover the choices that a generator could use to generate a given value.
Other work uses specifications to verify and synthesize generators.
Coverage types~\cite{zhou2023covering} describe values that a generator is guaranteed to generate.
Cobb~\cite{lafontaine2025covered} uses these types to repair generators that cannot generate all the specified values.
Clouseau~\cite{zhou2026clouseau} uses underapproximate trace specifications to synthesize effectful test generators.
\SystemName uses semantic choices as an executable witness that the modified generator can generate a fixed SQL target.

\section{Conclusion}

Extending a DBMS test generator requires implementing new SQL features and using them in queries that its oracle can check.
We have presented \SystemName, which uses coding agents to modify the generator and witness replay to check that it can generate target statements.
Generation witnesses record the database setup and choices needed to generate a target statement.
Replay checks these choices and reports failures to guide further source modifications.
We implemented \SystemName on top of \SQLancerPP so that the resulting generator can continue testing without model calls.
We evaluate feature acquisition separately from the bug-finding effectiveness of the evolved generators.

\bibliographystyle{ACM-Reference-Format}
\bibliography{sample-base}

\end{document}